\documentclass{ws-procs9x6}
\usepackage{amsmath,amssymb}

\def\x{\mbox{\boldmath$\displaystyle\mathbf{x}$}}
\def\p{\mbox{\boldmath$\displaystyle\mathbf{p}$}}
\def\r{\mbox{\boldmath$\displaystyle\mathbf{r}$}}
\def\0{\mbox{\boldmath$\displaystyle\mathbf{0}$}}
\def\s{\mbox{\boldmath$\displaystyle\mathbf{\sigma}$}}
\def\beq{\begin{equation}}
\def\eeq{\end{equation}}
\def\bv{\mbox{\boldmath$\displaystyle\mathbf{\varphi}$}}
\def\ba{\begin{array}}
\def\ea{\end{array}}

\begin{document}

\title{DARK MATTER AND DARK GAUGE FIELDS}

\author{D.~V.~AHLUWALIA}

\address{Department of Physics and Astronomy, Rutherford Building\\
University of Canterbury, Private Bag 4800\\ 
Christchurch 8020, New Zealand \\
E-mail: dharamvir.ahluwalia-khalilova@canterbury.ac.nz\\
www2.phys.canterbury.ac.nz/editorial/}

\author{CHENG-YANG~LEE, D.~SCHRITT, 
and T.~F.~WATSON}

\address{Department of Physics and Astronomy, Rutherford Building\\
University of Canterbury, Private Bag 4800\\ 
Christchurch 8020, New Zealand}

\begin{abstract}
Following the unexpected theoretical 
discovery of a mass dimension one fermionic quantum field of spin one half, we
now present first results on two {\em local} versions. The Dirac 
and Majorana fields of the standard model of particle physics are supplemented
by their natural counterparts in the dark matter sector. The possibility that
a mass dimension transmuting symmetry may underlie a new standard model 
of particle physics is briefly suggested.
\end{abstract}


\bodymatter

\section{Introduction: Dark matter and its darkness}
The existence of dark matter hints towards new physics,
and has often been thought to bring in 
new symmetries, such as super-symmetry. Yet, such
directions can also take us far afield if we have not
fully understood the technical and physical content of the
well-known continuous and discrete symmetries 
of the standard model of particle physics~\cite{Wigner1962:ep}.  
It is to this 
possibility that this paper is devoted in the most
conservative tradition of our craft.

At present, we do not know the spin of dark matter particles, 
nor do we have any inkling as to what type of 
interaction-inducing principle, if any, operates in the dark sector.
A highly motivated candidate that answers  these questions 
can suggest experiments that help make properties of dark matter
concrete. On the positive side, it is now well established that dark matter 
interacts with the particles of the standard model predominantly 
via gravity. All other interactions seem to be highly, if not
 completely, suppressed.

Without calling upon any yet-unobserved symmetries, we invoke the
minimal assumption that whatever dark matter is, it must transform in
a well-defined manner from one inertial frame to another according to
projective irreducible representations of the Poincar\'e algebra,
supplemented by the discrete symmetries of spacetime reflections and
charge conjugation. With this conservative ansatz we exploit recently
gained
insights~\cite{AhluwaliaKhalilova:2004sz,AhluwaliaKhalilova:2004ab} to
construct two fermionic spin one half quantum fields. These carry mass
dimension one, and satisfy the canonical locality requirement.

Defining darkness as the property
that one set of fields carries limited or no interactions (except
gravitational) with another set of fields, these new matter fields
 are dark with respect to the matter and gauge fields of the
 standard model (SM) of particle physics; and vice versa. 

Identifying the new fields with dark matter (DM), we suggest a 
mass-dimension-transmuting symmetry principle that yields the 
DM sector Lagrangian density without any additional assumptions.
This then becomes reminiscent of the super-symmetric paradigm but now
the mass-dimension-transmuting symmetry does not take a fermion to
a boson, but a SM fermion of mass dimension three-half to a DM 
fermion of mass dimension one.

\section{Two new quantum fields}

Here we present all the essential details 
for the construction
of two spin one half quantum fields based upon the  
dual helicity 
eigenspinors of the
charge conjugation operator (Elko)\footnote{The acronym {\em Elko} for the
eigenspinors of the charge conjugation operator 
originates from the 
German {\bf E}igenspinoren des 
{\bf L}adungs{\bf k}onjugations{\bf o}perators. 
The reader who wishes to study Elko similarities to and differences
from the Majorana spinors may wish to consult 
references~[\refcite{AhluwaliaKhalilova:2004ab,daRocha:2005ti}].}
introduced in 
references~[\refcite{AhluwaliaKhalilova:2004sz,AhluwaliaKhalilova:2004ab}].
The breakthrough that this paper presents, and what constitutes the 
primary progress 
 since the indicated 2005 publications,
is that the quantum fields we now present are {\em local} in the 
canonical sense.

\vspace{11pt}\noindent
\textit{Background\textemdash~}
To establish the notation and the working context, we here collect together
certain facts and definitions. In case the reader feels a certain
element of unfamiliarity, we urge her/him to
consult  Ref.~[\refcite{Ryder:1985wq,AhluwaliaKhalilova:2004ab}] for pedagogic details.

\vspace{11pt}
In the momentum space, dual helicity spinors have the generic form

\begin{equation}
\lambda(\p) = \left(
\begin{array}{c}
\left(\alpha \Theta\right)\phi_L^\ast(\p) \\
\phi_L(\p)
\end{array}
\right) 
\end{equation}
where $\Theta$ is the Wigner time reversal operator for spin one half
and reads

\begin{equation}
\Theta = \left(
\begin{array}{rr}
0 &  -1 \\
1 &  0
\end{array}\right).
\end{equation}
The $\phi_L(\p)$ is a massive left-handed (L) Weyl spinor, while
$\alpha \Theta \phi_L^\ast(p)$ transforms as a right-handed (R)
Weyl spinor associated with the same mass.
If $\phi_L(\p)$ carries a given helicity, then $\alpha \Theta \phi_L^\ast(p)$
is necessarily endowed with the opposite helicity.
The charge conjugation operator $C$ for the L$\oplus$R 
representation space is

\beq
C = \left(
\begin{array}{ccc}
\mathbb{O} & \empty~ &i \Theta \\
-i \Theta & \empty~ &\mathbb{O}
\end{array}
\right) \;K
\eeq
where $K$ is the  complex conjugation operator (see Ref.~[\refcite{AhluwaliaKhalilova:2004ab}]).
The $\lambda(\p)$ become Elko satisfying, $C\lambda(\p) = \pm \lambda(\p)$, if
$\alpha = \pm i$.
To construct a complete set of Elko we need the 
L$\oplus$R boost operator, and a complete 
dual-helicity basis for Elko at rest; i.e., $\lambda(\0)$. 
The boost operator reads

\beq
\kappa = 
\left(\begin{array}{cc}
\exp\left(\frac{\s}{2}\cdot\bv\right) & \mathbb{O} \\
 \mathbb{O} & \exp\left(- \frac{\s}{2}\cdot\bv\right)
\end{array}
\right).
\eeq
In terms of the energy $E$ and the momentum $\p = p \hat\p$ the 
boost parameter, $\bv = \varphi \hat\p$, is defined
as $\cosh(\varphi) = E/m$, and $\sinh(\varphi) = p/m$, where
$m$ is the mass of the described particle. Since 
$\left(\s\cdot\hat\p\right)^2 = 
\mathbb{I}$, the boost operator is linear in $\p$\footnote{It is
this fact, when coupled with the observation that the Dirac spinors
are eigenspinors of the L$\oplus$R  parity operator,
that yields the linearity of the Dirac operator $\left(
i \gamma^\mu \partial_\mu - m \right)$. For a complementary
discussion, the reader is referred to Sec. 5.5 of reference 
[\refcite{Weinberg:1995mt}].}

\beq
\kappa= \sqrt{\frac{E+m}{2 m}}\left(
\begin{array}{cc}
\mathbb{I} + \frac{\s\cdot\p}{E+m} & \mathbb{O} \\
\mathbb{O} & \mathbb{I} - \frac{\s\cdot\p}{E+m}
\end{array}
\right).
\eeq

\vspace{11pt}\noindent
\textit{Construction of Elko for the two new quantum fields\textemdash~}
To construct the required Elko, we must first introduce a basis for the
particle description in its rest frame. Towards this end we shall take
$\phi_L(\0)$ to be eigenspinors of the helicity operator $(\s/2)\cdot\hat\p$

\beq
\left[\frac{\s}{2}\cdot\hat\p\right] \phi_L^\pm(\0) = \pm \frac{1}{2} \phi_L^\pm(\0).
\eeq
Taking $\hat \p = \left(\sin\theta\cos\phi,\sin\theta\sin\phi,\cos\theta\right)$,
we adopt a certain choice of phase factors so that\footnote{Please notice that
the choice of phases here differs from that given in 
references~[\refcite{AhluwaliaKhalilova:2004sz,AhluwaliaKhalilova:2004ab}].} 

\beq
\phi^+_L(\0) = \sqrt{m} \left(\ba{c}
\cos(\theta/2) e^{- i \phi/2}\\
\sin(\theta/2) e^{+i \phi/2}
\ea
\right),\;
\phi^-_L(\0) = \sqrt{m} \left(\ba{c}
-\sin(\theta/2) e^{- i \phi/2}\\
\cos(\theta/2) e^{+i \phi/2}
\ea
\right).
\eeq 
The complete basis for Elko that is now required
 for constructing 
the indicated quantum fields is

\begin{eqnarray}
&&\xi_{\{-,+\}}(\0) = 
\lambda(\0)\Big\vert_{\phi_L(\mathbf{0})\rightarrow 
\phi^+_L(\mathbf{0}),\;\alpha\rightarrow + i}\label{eq:a}\\
&&\xi_{\{+,-\}}(\0) = 
\lambda(\0)\Big\vert_{\phi_L(\mathbf{0})\rightarrow 
\phi^-_L(\mathbf{0}),\;\alpha\rightarrow + i} \label{eq:b}\\
&&\zeta_{\{-,+\}}(\0) = 
\lambda(\0)\Big\vert_{\phi_L(\mathbf{0})\rightarrow 
\phi^-_L(\mathbf{0}),\;\alpha\rightarrow - i}\label{eq:c}\\
&&\zeta_{\{+,-\}}(\0) = -
\lambda(\0)\Big\vert_{\phi_L(\mathbf{0})\rightarrow 
\phi^+_L(\mathbf{0}),\;\alpha\rightarrow - i} \label{eq:d}
\end{eqnarray}
with

\beq
\xi_{\{\mp,\pm\}}(\p) = \kappa\, \xi_{\{\mp,\pm\}}(\0),\quad
 \zeta_{\{\mp,\pm\}}(\p) = \kappa\, \zeta_{\{\mp,\pm\}}(\0).
\eeq
The justification
for the choice of phases and designations in equations
(\ref{eq:a}-\ref{eq:d}) is far from trivial; but it
is a straightforward generalisation of the  
reasoning found in Sec. 38 of 
reference~[\refcite{Srednicki:2007qs}] and Sec. 5.5 of 
reference~[\refcite{Weinberg:1995mt}].

If one works with Elko using the Dirac dual, $\overline\eta(\p) = 
\eta^\dagger(\p) \gamma^0$, one finds\footnote{Here $\eta$ stands 
symbolically for either $\xi$ or $\zeta$.} that these 
carry null norm\cite{Ahluwalia:1994uy}. One also encounters problems 
such as those found in Appendix P of reference~[\refcite{Aitchison:2004cs}].
In fact, the dual-helicity nature of the $\xi(\p)$ and $\zeta(\p)$ asks for the
introduction of a new dual. We shall call it the
{\em Elko dual}.\footnote{The origin of this dual 
dates back to a preliminary work~[\refcite{AhluwaliaKhalilova:2003jt}]
on the subject.}
 It is defined as

\beq
 \stackrel{\neg}\eta_{\{\mp,\pm\}}(\p) := \mp\, i\, \eta_{\{\pm,\mp\}}^\dagger(\p)\,
\gamma^0
\eeq 
where

\beq
\gamma^0:= \left(\ba{cc}
\mathbb{O} & \mathbb{I} \\
\mathbb{I} & \mathbb{O}
\ea\right).
\eeq
With the Elko dual thus defined, we now have, by construction, the {\em orthonormality 
relations}

\begin{eqnarray}
&&  \stackrel{\neg}\xi_\beta(\p) \, \xi_{\beta^\prime}(\p) = 
+ \,2 m \, \delta_{\beta\beta^\prime}\;,\quad
 \stackrel{\neg}\xi_\beta(\p) \, \zeta_{\beta^\prime}(\p) = 
0   \\
&& \stackrel{\neg}\zeta_\beta(\p) \, \zeta_{\beta^\prime}(\p) = 
-\, 2 m \, \delta_{\beta\beta^\prime}\;, \quad
  \stackrel{\neg}\zeta_\beta(\p) \, \xi_{\beta^\prime}(\p) = 
0 
\end{eqnarray}
Here, $\beta$ ranges over two possibilities: $\{+,-\}$ and  $\{-,+\}$.
The {\em completeness relation} is 

\beq
\frac{1}{2 m}\sum_\beta
\Big[\xi_\beta(\p)\,\stackrel{\neg}\xi_\beta(\p)
- \zeta_\beta(\p)\,\stackrel{\neg}\zeta_\beta(\p)\Big] = \mathbb{I}.
\eeq
The detailed structure underlying the completeness relations resides
in the following {\em spin sums}

\begin{eqnarray}
\xi_\beta(\p)\,\stackrel{\neg}\xi_\beta(\p) 
= + m \left[\mathbb{I} + \mathcal{G}(\p)\right] \label{eq:spinsumxi}\\
\zeta_\beta(\p)\,\stackrel{\neg}\zeta_\beta(\p) 
= - m \left[\mathbb{I} - \mathcal{G}(\p)\right]  \label{eq:spinsumzeta}
\end{eqnarray}
which together {\em define} $\mathcal{G}(\p)$. Explicit
calculation shows that $\mathcal{G}$ is an odd function of $\p$

\beq
\mathcal{G}(\p) = - \mathcal{G}(-\p). \label{eq:Godd}
\eeq
It must be noted at this stage that the entire set of results
is specific to the helicity basis. A
careful examination is needed if one wishes to consider
a basis in which  $\phi_L(\p)$ are not eigenspinors of
the helicity operator but, say, some other operator of the type 
 $(\s/2)\cdot\hat \r$; where $\hat \r$  is not coincident
with $\hat\p$. These two bases, as is evident on reflection, 
represent two {\em physically distinct} situations. Our preliminary
results confirm that additional work is needed to extract further
physical and mathematical content in such situations.

\vspace{11pt}\noindent
\textit{Two new quantum fields based on Elko: Erebus and Nyx\textemdash~}
Following Greek mythology about primordial darkness 
we now introduce two fields which will
later be identified with dark matter. We define Erebus as

\beq
e(x) = \int \frac{d^3 p}{(2\pi)^3}\frac{1}{\sqrt{2 m E(\p)}} 
\sum_\beta \left[ c_\beta(\p) \xi_\beta(\p) e^{-i p_\mu x^\mu} 
+  d^\dagger_\beta(\p) \zeta_\beta(\p) e^{+i p_\mu x^\mu}\right].\nonumber
\eeq
It enjoys the same formal status as the Dirac field constructed from
the Dirac spinors. We define Nyx as 

\beq
n(x) = \int \frac{d^3 p}{(2\pi)^3}\frac{1}{\sqrt{2 m E(\p)}} 
\sum_\beta \left[ c_\beta(\p) \xi_\beta(\p) e^{-i p_\mu x^\mu} 
+  c^\dagger_\beta(\p) \zeta_\beta(\p) e^{+i p_\mu x^\mu}\right].\nonumber
\eeq
It enjoys at the same formal status as the Majorana field constructed from
the Dirac spinors. These formal similarities betray the true physical
content of these new fields as we shall discover on deeper analysis.
In a nutshell the difference is this:  the mass dimension of
Erebus and Nyx is one, not three half (see below). 
It is this aspect that renders
them  dark with respect to the Dirac and Majorana fields of the
standard model of particle physics.\footnote{The reader may wish to
consult reference~[\refcite{AhluwaliaKhalilova:2004ab}] for the details of
this argument.}
The creation, $c^\dagger_\beta$ and $d^\dagger_\beta$, 
and annihilation, $c_\beta$ and $d_\beta$, operators 
that appear in the $e(x)$
and $n(x)$ fields satisfy the standard fermionic anti-commutation 
relations

\begin{eqnarray}
\left\{c_\beta(\p),\;c^\dagger_{\beta^\prime}(\p^\prime)\right\} =
(2\pi)^3\, \delta_{\beta\beta^\prime} \,\delta^3(\p-\p^\prime) \\
\left\{c_\beta(\p),\;c_{\beta^\prime}(\p^\prime)\right\} =
\left\{c^\dagger_\beta(\p),\;c^\dagger_{\beta^\prime}(\p^\prime)\right\} =0
\end{eqnarray}
with similar relations for the $d$'s.

The Elko duals of Erebus and Nyx fields, $\stackrel{\neg}e(x)$ and 
$\stackrel{\neg}n(x)$, are obtained by the following substitutions in
$e(x)$ and $n(x)$

\begin{eqnarray}
&& \xi_\beta(\p) \rightarrow \stackrel{\neg}\xi_\beta(\p),\quad
\zeta_\beta(\p) \rightarrow \stackrel{\neg}\zeta_\beta(\p) \nonumber\\ 
&& e^{\pm i p_\mu x^\nu} \leftrightarrow e^{\mp i p_\mu x^\nu},\quad
 c_\beta(\p) \leftrightarrow c^\dagger_\beta(\p),\quad
 d_\beta(\p) \leftrightarrow d^\dagger_\beta(\p).
\end{eqnarray}
The propagator associated with the Erebus and Nyx fields
follows from textbook methods, and is further elaborated 
in reference~[\refcite{AhluwaliaKhalilova:2004ab}]. It entails 
evaluation of $\langle~\vert T(e(x^\prime) \stackrel{\neg}e(x)\vert~\rangle$,
and  $\langle~\vert T(n(x^\prime) \stackrel{\neg}n(x)\vert~\rangle$,
where $T$ is the fermionic time-ordering operator, and $\vert~\rangle$ 
represents the vacuum state. The result for Erebus as well as Nyx,
in terms of spin sums, reads

\begin{eqnarray}
S(x-x^\prime) = - \int \frac{d^3 p}{(2\pi)^3}&& \frac{i}{2 m E(\p)}
 \sum_\beta {\Big[}\theta(t^\prime - t) \,\xi_\beta(\p) 
{\stackrel{\neg}\xi}_\beta(\p) 
e^{- i p_\mu(x^{\prime\mu} - x^\mu)}\nonumber \\
&& - \theta(t - t^\prime) \,\zeta_\beta(\p) 
{\stackrel{\neg}\zeta}_\beta(\p) 
e^{+i p_\mu(x^{\prime\mu} - x^\mu)} {\Big]}
\end{eqnarray}
Using the spins sums~(\ref{eq:spinsumxi}) and
(\ref{eq:spinsumzeta}) yields (again, see reference~\refcite{AhluwaliaKhalilova:2004ab} for additional details)

\beq
S(x-x^\prime) = \int\frac{d^4 p}{(2\pi)^4} e^{i p_\mu (x^\mu - x^{\prime\mu})}
\frac{\mathbb{I} + \mathcal{G(\p)}}{p_\mu p^\mu - m^2 + i\epsilon}.
\eeq
Here, the limit $\epsilon\to 0^+$ is implicit. 
In view of (\ref{eq:Godd}),
it is clear that, in the absence of a preferred direction, 
such as one arising from an external magnetic field, the second term in 
the above equation identically vanishes. As a result, 
$S(x-x^\prime)$ reduces to the Klein-Gordon propagator, modulo a $4\times 4$
multiplicative identity matrix in the R$\oplus$L representation space.
Consequently, both the Erebus and Nyx fields carry mass dimension one. 
This mass dimensionality forbids particles described by these fields to enter $SU(2)_L$ 
doublets of the SM. The fields $e(x)$ and $n(x)$ thus become  first-principle
candidates for dark matter. Following arguments similar to those
presented in~[\refcite{AhluwaliaKhalilova:2004sz,AhluwaliaKhalilova:2004ab}]
the free Lagrangian densities associated with the Erebus and Nyx fields
are

\begin{eqnarray}
\mathcal{L}^{\mathrm{Erebus}} = \partial^\mu{\stackrel{\neg}e}(x)
\partial_\mu e(x) - m_e^2 {\stackrel{\neg}e}(x)  e(x)\label{eq:Le}\\
\mathcal{L}^{\mathrm{Nyx}} = \partial^\mu{\stackrel{\neg}n}(x)
\partial_\mu n(x) - m_n^2 {\stackrel{\neg}n}(x)  n(x) \label{eq:Ln}
\end{eqnarray}
Here, we have made it explicit that the Erebus and Nyx fields
need not carry the same mass.

\section{A Pause}

A brief pause in the development of 
our presentation now appears
necessary. We physicists are trained to consider Lagrangian density, 
$\mathcal{L}$, almost
as a God-given entity. Granted, one places some restrictions such as that
it must be a Lorentz scalar, but beyond these well known caveats that
$\mathcal{L}$ rarely is anything but a product of one's genius (such
as was the case in Dirac's 1928 paper~[\refcite{Dirac:1928hu}]) 
or a well-educated guess
which often times runs along the lines, ``for simplicity let's
assume that $\mathcal{L}$ is linear in derivatives, ...''. This
is all wrong, in essence.
 Or, so is the lesson that stares us
in the face of above-derived $\mathcal{L}$ for Erebus and Nyx. 
In fact, one could have as easily derived  $\mathcal{L}^{\mathrm{Dirac}}$
had one started not with the Elko, but with eigenspinors of the parity
operator in the R$\oplus$L representation space.\footnote{Please note
that the usual parity operator in the R$\oplus$L representation space
can be easily constructed without reference to 
 $\mathcal{L}^{\mathrm{Dirac}}$, or the  $\mathcal{L}$ for the
Erebus and Nyx fields. If this is not obvious, kindly wait for
one of the sequels to this paper.}

In a nutshell: The kinematic structure as contained in
 $\mathcal{L}$ is determined entirely once one agrees that
it must be based on a quantum field, and that this quantum field
has certain properties under the symmetry operations inherent
to the considered representation space. The details are obvious from
the presented example of Erebus and Nyx.

\section{Locality structure of the Erebus and Nyx fields}

Having arrived at 
 $\mathcal{L}$ for Erebus and Nyx in (\ref{eq:Le}) and 
(\ref{eq:Ln}) one can immediately obtain the canonically conjugate momenta
$\pi(x)$ for these fields. That done, a straightforward, though slightly 
lengthy calculation involving various Elko `spin sums' 
 shows that Erebus and Nyx are local. For Erebus we find 

\begin{eqnarray}
&& \left\{e(\x,t), \pi(\x^\prime,t)\right\} = i \delta^3(\x-\x^\prime) \\
&& \left\{e(\x,t), e(\x^\prime,t)\right\} = 0,\;
 \left\{\pi(\x,t), \pi(\x^\prime,t)\right\} =0
\end{eqnarray}
An exactly similar set of equal-time anti-commutators exist for Nyx.

These results are in remarkable contrast, and constitute the 
breakthrough alluded to above,
to the non-locality structure presented in 
references~[\refcite{AhluwaliaKhalilova:2004sz,AhluwaliaKhalilova:2004ab}]. 
These results also provide a counter example to the expectations
based on the 1966 work of Lee and Wick\cite{Lee:1966ik}. A resolution of
this apparent discrepancy remains to be found.

\section{Towards a standard model with Erebus and Nyx fields}

An intriguing fact that emerges from the above analysis is that
the CPT structure of the spin one half R$\oplus$L representation space is far
richer than one would have expected. 
Such an analysis proceeds along similar lines as found in
references~[\refcite{AhluwaliaKhalilova:2004sz,AhluwaliaKhalilova:2004ab}]
and yields, e.g., $(CPT)^2 = - \mathbb{I}$ for the Erebus and Nyx fields.
The Dirac and Majorana fields
are endowed with mass dimension and CPT properties that are
in sharp contrast to those of  the Erebus and Nyx fields.
Therein may lie novel sources of violation of discrete symmetries, 
such as parity and combined symmetry of charge conjugation and parity,
and these may provide insights into such questions as raised by 
Saunders\cite{Saunders:2007parity}.

In the present paper we do not further analyse these matters, 
but instead pose the question: does there exist a 
fundamental principle that one may invoke to extend the standard model
of particle physics to systematically incorporate the Erebus and Nyx fields?
We conjecture a symmetry operator $\mathcal{A}$ 
such that\footnote{A related work has just appeared on the subject.
We refer the reader to reference~[\refcite{daRocha:2007pz}] 
for details.}

\begin{equation}
\mathcal{A} \mathcal{L}^{\mathrm{Dirac}} \mathcal{A}^{-1}  \rightarrow
 \mathcal{L}^{\mathrm{Erebus}}
\end{equation}
or, equivalently,
$\mathcal{A}^{-1} \mathcal{L}^{\mathrm{Erebus}} \mathcal{A}  \rightarrow
 \mathcal{L}^{\mathrm{Dirac}}$.\footnote{With a similar symmetry 
existing between the Majorana and Nyx fields.} Clearly, such a symmetry,
if it exists,
transmutes the mass dimension of spin one half fermionic fields from
3/2 to 1 and from 1 to 3/2. The principle of local gauge interactions 
for the Erebus and Nyx fields
in its $U(1)$ form is different from that of the SM: it corresponds to
invariance of the Lagrangian density
under $\vert e \rangle \rightarrow \exp[i \gamma^5
\alpha(x)]\, \vert e \rangle$, and $\vert n \rangle \rightarrow \exp[i \gamma^5
\alpha(x)]\, \vert n \rangle$. For the dark gauge fields we choose the 
name Shakti. This time we have invoked the mythology of the East.

Should $\mathcal{A} $ exist, and be realised by
nature, the extended Lagrangian density would form a first-principle
extension of the standard model to incorporate dark matter; i.e., 
if Erebus and Nyx indeed describe dark matter. The latter 
carries high plausibility  given the natural darkness of the $e(x)$ and 
$n(x)$ fields with respect to the SM fields. We hasten to add that
the Erebus and Nyx, just like their SM counterparts, are not, 
or need not, be self-referentially dark. It is prevented by Shakti.

\section{Concluding remarks}

We have presented two new local quantum fields based on the
dual helicity eigenspinors of the charge conjugation operator. 
These are fermionic, and carry mass dimension one. The latter
result provides an {\em ab initio} origin of darkness 
of dark matter, and at the same time suggests the possible 
existence of a mass dimension transmuting symmetry which
may be used to construct an extension of the 
standard model of particle physics to include the dark matter sector. 

The SM matter and gauge fields and their counterparts
in Erebus, Nyx, and Shakti  
may provide two self-referentially luminous sectors of the cosmos 
which, although 
dark  to each other, may be united by Higgs and gravity.
Indeed, one may imagine whispering 
dark rivers and silvery dark moons in the night 
of the world inhabited by Erebus, Nyx, and Shakti \textemdash~ a world we
see not, and yet a world to which owe our existence.

\section*{Acknowledgements} 

It is our great pleasure to thank
all the organisers of ``Dark 2007 -- 
Sixth International Heidelberg Conference on Dark Matter 
in Astro and Particle Physics'' for their warm hospitality.

On a personal note we are grateful to two dedicated
powerhouses of physics,
Hans  Klapdor-Kleingrothaus and Irina Krivosheina, who 
by their relentless pursuit of knowledge and their scholarship
have added to physics and its culture the spirit so well captured
in Herman Hesse's {\em Das Glasperlenspiel} (The Glass Bead Game).
This work was inspired by a talk Hans gave more than
a decade ago at the Los Alamos National Laboratory and which led the 
senior author of the present manuscript 
to investigate Majorana's 1937 
paper\cite{Majorana:1937vz,KlapdorKleingrothaus:2001ke,KlapdorKleingrothaus:2006ff}. That in
a series of papers Hans and his collaborators have provided a
first glimpse into this beautiful realm attests to the genius
of his single-minded pursuits and unique abilities. We give Hans, Irina and
their team our very best wishes.

\end{document}